\documentclass[aps,prd,twocolumn,superscriptaddress]{revtex4}
\usepackage{graphicx}% Include figure files
\usepackage{amsmath}
\usepackage{amssymb}
\usepackage{slashed}
\usepackage{latexsym}
\usepackage{epsfig}
\usepackage{amsbsy}
\usepackage{array}
\usepackage{amssymb}
\usepackage{setspace}
\usepackage{bm}
\usepackage{lipsum}
\usepackage{mathrsfs}
\usepackage{float}
\usepackage{color}
\usepackage[T1]{fontenc}
\usepackage{mathptmx}
\DeclareMathAlphabet{\mathcal}{OMS}{cmsy}{m}{n}
\DeclareSymbolFont{largesymbols}{OMX}{cmex}{m}{n}

\begin{document}

\title{Do the current astronomical observations exclude the existence of non-strange quark stars?}

\author{Tong Zhao}
\email{zhao708@purdue.edu}
\affiliation{Department of physics, Nanjing University, Nanjing 210093, China}
\author{Wei Zheng}
\email{007128@yzu.edu.cn}
\affiliation{College of Physics Science and Technology, Yangzhou University, Yangzhou 225009, China}
\author{Fei Wang}
\email{1194899400@qq.com}
\affiliation{Department of Physics, Nanjing University, Nanjing 210093, China}
\author{Cheng-Ming Li}
\email{licm.phys@gmail.com}
\affiliation{Department of physics, Nanjing University, Nanjing 210093, China}
\author{Yan Yan}
\email{2919ywhhxh@163.com}
\affiliation{School of mathematics and physics, Changzhou University, Changzhou, Jiangsu 213164, China}
\author{Yong-Feng Huang}
\email{hyf@nju.edu.cn}
\affiliation{Department of Astronomy, Nanjing University, Nanjing 210093, China}
\author{Hong-Shi Zong}
\email{zonghs@nju.edu.cn}
\affiliation{Department of Physics, Nanjing University, Nanjing 210093, China}
\affiliation{Joint Center for Particle, Nuclear Physics and Cosmology, Nanjing 210093, China}
\affiliation{State Key Laboratory of Theoretical Physics, Institute of Theoretical Physics, CAS, Beijing 100190, China}

\begin{abstract}
   As is pointed out in a recent work (Phys. Rev. Lett. 120, 222001), quark matter may not be strange. Inspired by this argument, we use a new self-consistent mean field approximation method to study the equation of state of cold dense matter within the framework of the two-flavor NJL model. Then the mass-radius relationship of two-flavor pure quark stars is studied. In the new self-consistent mean field approximation method we introduced, there is a free parameter $\alpha$, which reflects the weight of different interaction channels. In principle, $\alpha$ should be determined by experiments rather than the self-consistent mean field approximation itself. In this paper, thus, the influence of the variation of $\alpha$ on the critical chemical potential of chiral phase transition and the stiffness of the EOS are thoroughly explored. The stiffness of the EOS can be increased greatly to support a two-solar-mass pure quark star when $\alpha$ is greater than 0.9, because the contribution from the vector term is retained by the Fierz transformation. Our result is also within the constraints on the radius from the recent data analysis of the tidal deformability. This means that the current theoretical calculations and astronomical observations cannot rule out the possibility of a two-flavor pure quark star.

%\pacs{12.38.Aw, 12.39.Ba, 14.65.Bt, 97.60.Jd}

\end{abstract}

\maketitle
\section{Introduction}
The equation of state (EOS) is the crux of the studies of neutron stars. Once the EOS describing a single or several phases is specified, the mass-radius relation can be obtained by integrating the Tolman-Oppernheimer-Volkoff (TOV) equation. Recent results derived from observations have provided serious constraints on EOSs. On one hand, there has been strong evidence that 2-solar-mass neutron stars exist \cite{1,2,3,mrnew}. On the other hand, the gravitational-wave signal from the neutron star merger GW170817 provides not only a constraint on the tidal deformability but also possible constraints on neutron star radii \cite{4,5,6,7,8}. However, both the mass-radius relation and the inner structure of neutron stars are severely model-dependent.

It is generally believed that the description of the matter in term of interacting nucleons is valid when the density of matter is smaller than two times of nuclear saturation density $n_0$. Then, in the region of $2n_0 \sim 7 n_0$, the system gradually changes from hadronic to quark matter. When the density is greater than $4 n_0$, the matter is percolated and quarks no longer belong to specific baryons \cite{9}. So, neutron stars are usually considered as hybrid stars and the EOSs are also hybrid containing both nuclear matter, deconfined quark matter, and even mixed matter. In the conventional description of the onset of quark matter, hadronic matter (hyperons can also be included) and quark matter are two distinct phases. Then, the Maxwell construction is conceived to guarantee pressure and chemical potential continuity across the transition \cite{10,11,12,13}. In recent studies, between the hadronic matter and quark matter, a smooth crossover and a so-called quarkyonic regime at the intermediate baryon density are introduced to remedy our limited understanding of the hadron-quark phase transition \cite{14,15}.

In this work, we want to point out that there is a subtle paradox in most of hybrid star models. From the picture of quark degrees of freedom, the strong interaction matter will undergo the well-known chiral phase transition along with the continuous increase of the quark chemical potential to a critical value $\mu _c$. Because Lattice Quantum Chromodynamics (QCD) is invalid to deal with the large chemical potential problem at present, the value of $\mu _c$ depends on the phenomenological QCD models. The value of $\mu _c$ is predicted to be about 330-380 MeV in most Nambu-Jona-Lasinio (NJL) type models \cite{16,17,18,add1,add6,add5}. Other phenomenological QCD models from the quark degrees of freedom also give a similar prediction \cite{add7,19,add2,add3,add4}. Nonetheless, from the picture of hadron degrees of freedom, the quark chemical potential corresponding to the $4n_0$ is around 600 MeV. Even $2n_0$ indicates a quark chemical potential around 400 MeV \cite{20,21,22,23}. A recent work try to constrain the Hadron-Quark phase transition chemical potential via astronomical observations \cite{constrain}. According to the Ref. \cite{constrain}, the most possible baryon chemical potential where hadron matter disappears totally is 1.49 GeV corresponding to a 500 MeV quark chemical potential. All of these facts indicate that there is a huge contradiction between the model predictions derived from quark degrees of freedom and hadron degrees of freedom. To resolve this contradiction, a new approach of self-consistent mean field approximation is proposed for NJL type models in our previous work \cite{20}. In this paper, we derive the EOS at finite density based on our new self-consistent mean field approximation approach, and integrate the TOV equation to calculate the mass-radii relation of quark stars.

There are two motivations for doing this. First, although stable quark stars can exist based on the theory of E. Witten where strange quark matter might be the ground state of strong-interaction matter \cite{24,25}. With the lack of a first-principles understanding of the strong interaction at finite density, the MIT bag model was used in early works \cite{26}. Since then, most works of the inner structure of neutron stars have assumed that the neutron stars are composed of $u,d$ and $s$ three-flavor quarks. For a recent example, see Ref. \cite{lbl}. %The MIT bag model treats quarks as free fermions in a bag with negative bag pressure. And the energy per baryon of quark in this model is quite sensitive to the assumed QCD parameters.
However, a recent study shows that stable quark matter may not be strange, and two-flavor quark stars with a larger maximum mass can also exist \cite{28}. It should be noticed here, at present, it is an open question whether the most stable matter is two-flavor or three-flavor quark matter. So, we want to seek for the possibility that a two-solar-mass quark star contains only $u$ and $d$ quarks in this paper. Second, we recently proposed a new self-consistent mean field approximation method by means of Fierz transformation \cite{20}, and applied it to the two-flavor NJL model. This new self-consistent mean field approximation introduces a new free parameter $\alpha$ that reflects the weight of different interaction channels. The parameter $\alpha$ in our model can influence not only the value of the critical chemical potential $\mu _c$ of chiral phase transition, but also the continuity of the chiral susceptibility. The chiral phase transition can become a crossover if $\alpha$ is greater than 0.71. So, it is interesting to investigate how it will influence the maximum mass of compact stars predicted by our model.

This paper is organized as follows: In Sec. II, we introduce our self-consistent mean field theory of the NJL model. In Sec. III, the mass-radii relations are calculated by solving the TOV equation and the effects of the parameters on the stiffness of the EOS are explored. Sec. IV is the summary and discussion of our work.

\section{Self-consistent mean field approximation}
The NJL model is widely adopted as a quark model to describe cold dense matter in neutron stars and quark stars \cite{29}. The standard  two-flavor NJL Lagrangian with interaction terms in the scalar and pseudoscalar-isovector channels is given by:
\begin{eqnarray}
\mathcal{L}=\bar{\psi }(i\slashed{\partial } - m)\psi + G[\left(\psi \bar{\psi }\right)^2+\left(\bar{\psi } i \gamma ^5 \tau \psi \right)^2],
\end{eqnarray}
where m is the current quark mass and G denotes the coupling constant. The Fierz identity of it is:

\begin{eqnarray}
\begin{aligned}
\mathcal{L} _F=& \bar{\psi }(i\slashed{\partial } - m)\psi + G\frac{1} {8 N_c}[2\left(\bar{\psi } \psi \right)^2+2\left(\bar{\psi } i \gamma ^5 \tau \psi \right)^2\\
& - 2\left(\bar{\psi } \tau \psi \right)^2 - 2\left(\bar{\psi } i \gamma ^5 \psi \right)^2 - 4\left(\bar{\psi } \gamma^{\mu } \psi \right)^2\\
& - 4\left(\bar{\psi } \gamma^{\mu } \gamma^5\psi \right)^2 + \left(\bar{\psi } \sigma^{\mu \nu} \psi \right)^2 - \left(\bar{\psi } \sigma^{\mu \nu} \tau \psi \right)^2].
\end{aligned}
\end{eqnarray}
Comparing Eqs. (1) and (2), it is easy to find that the contribution of the next leading order term $O(1/N_c)$ of the large $N_c$ expansion can be introduced by the Fierz transformation in the framework of the mean-field approximation. Because Fierz transformation is a mathematically equivalent transformation, the original Lagrangian and the transformed Lagrangian are equivalent. One can also construct a refined new equivalent Lagrangian by taking the linear combination of them:

\begin{eqnarray}
\mathcal{L} _R=\left(1-\alpha \right)\mathcal{L} + \alpha \mathcal{L} _F ,
\end{eqnarray}
where $\alpha$ is an arbitrary complex number. Just as it is pointed out in Ref. \cite{29}, if the mean field approximation is applied here to study the position of the phase transition of strong interaction at finite chemical potential, mean field Lagrangian $\langle\mathcal{L} \rangle$ and $\langle\mathcal{L} _F \rangle$ will give different results. The Ref. \cite{29} suggests this form: $1/2\langle\mathcal{L} \rangle + 1/2\langle\mathcal{L} _F \rangle$ $\left(\alpha=1/2 \right)$ to include both the Hartree term and Fock term, which means the contributions of them are assumed to be identical. Nonetheless, it is obviously that there are no physical requirements for the value of $\alpha$. And actually $\alpha$ should be constrained by experiments. Currently, with the lack of experiment data on finite density strongly interacting matter, in our model $\alpha$ is assumed to be an arbitrary complex number and our redefined mean field Lagrangian is:
\begin{eqnarray}
\langle\mathcal{L} _R \rangle=\left(1-\alpha \right)\langle\mathcal{L} \rangle + \alpha\langle\mathcal{L} _F \rangle.
\end{eqnarray}

Indeed, the parameter $\alpha$ in our manuscript reflects the weight of vector-isoscalar channel contribution in the case of finite density. Just as is pointed out in Ref. \cite{Walecka}, the vector-isoscalar channel is very important at finite chemical potential. Considered this fact, the  vector-isoscalar term is added into the standard NJL model by hand in some model studies\cite{17}. However, it should be noted here that those model Lagrangian are different from that of the original standard NJL model. And the artificially introduced interaction items lead to more parameters, which causes the standard NJL model to lose its predictive power.For example, when the case of finite chemical potential is discussed, the vector-isoscalar channel is artificially added to the standard NJL model Lagrangian.  Similarly, if the axial chemical potential is discussed (in this case, isovector-isoscalar channel is very important), the isovector-isoscalar channel is also artificially added to the standard NJL model (for the introduction of axial chemical potential, see Ref. \cite{zong1}). More importantly, if the Fierz transformation is not considered, it means that only the contribution of the leading order term of the large $N_c$ expansion is considered and the next leading term $O(1/N_c)$ is ignored in the mean field approximation. As the authors of Ref. [41] pointed out, since the Fierz transformation is not considered, the mean field approximation approach at this time is theoretically not self-consistent.\\In contrast, our self-consistent mean field approximation approach avoids this arbitrariness and can be handled in a self-consistent manner for any background field, for example, in the case of strong magnetic field. Because there exists a strong magnetic field on the  surface of neutron stars \cite{Shapiro}, we must consider the effects of the strong magnetic field in the investigations of  the EOS of the neutron star.
%  It is known that, the phase transition is the result of competition between different interaction channels. In different environments, the weight of each interaction channel is different. } %Because the Fierz transformation reveals different interaction channels in the Lagrangian. After the mean field approximation is applied, both the vector channel $\left(\bar{\psi } \gamma^0 \psi \right)^2$ and the scalar channel $\left(\bar{\psi } \psi \right)^2$ are left in the Fierz transformed Lagrangian. However, there will be no vector channel contributions in the original Lagrangian if the mean-field approximation is applied.

Based on our new Lagranian (4), the two-flavor gap equation is then given by:

\begin{eqnarray}
\begin{aligned}
M_i &= m_i + \left(12-11\alpha \right)g \times \\
&\sum _ {f = u, d} \frac{M_f} {\pi^2}\int_ 0^{\Lambda}\frac {p^2} {E_{pf}}[1-n_{pf}\left(T,\mu _{rf} \right)-\bar{n} _{pf} \left(T,\mu _{rf} \right)] \, dp,
\end{aligned}
\end{eqnarray}
where $i=u$ or $d$, $m_u = m_d = 5~\mathrm {MeV}$ is the current quark mass, $E_{pi} =\sqrt {p^2 + M_i ^2}$, $N_c =3$, $g=\frac{G} {2-11\alpha /6}$ can be fixed by the coupling constant $G=4.93\times 10^{-6} \mathrm {MeV}^{-2}$, $\Lambda = 653\mathrm {MeV}$ is the three-momentum cutoff, $\mu _r$ is the effective chemical potential that satisfies:

\begin{eqnarray}
\begin{aligned}
\mu _{ri} =& \mu_i - \frac{6\alpha g} {N_c \pi^2}\times \\
&\sum _ {f = u, d} \int_ 0^{\Lambda} p^2 [n_{pf}\left(T,\mu _{rf} \right)-\bar{n} _{pf} \left(T,\mu _{rf} \right)] \, dp,
\end{aligned}
\end{eqnarray}
and

\begin{eqnarray}
\begin{aligned}
n_{pi}\left(T,\mu _{ri} \right)&=\frac {1} {\exp\left (\frac {E_{pi} - \mu _{ri}} {T} \right) + 1},\\
\bar{n} _{pi} \left(T,\mu _{ri} \right)&=\frac {1} {\exp\left (\frac {E_{pi} + \mu _{ri}} {T} \right) + 1}.
\end{aligned}
\end{eqnarray}
Then, the number density is given by:

\begin{eqnarray}
\rho_i= \frac{3} {\pi^2} \int_ 0^{\Lambda} p^2 [n_{pi}\left(T,\mu _{ri} \right)-\bar{n} _{pi} \left(T,\mu _{ri} \right)] \, dp.
\end{eqnarray}
To get the EOS for the asymmetric matter in quark stars, we take the chemical equilibrium into account. The chemical equilibrium and electric neutrality for the two-flavor quark matter are:

\begin{eqnarray}
\begin{cases}
  \mu _d &=\mu _u +\mu _e, \\
  \frac{2} {3} \rho _u &= \frac{1} {3} \rho _d  +\rho _e.
\end{cases}
\end{eqnarray}
Combining Eqs. (5-9), we get the relations between densities of u,d quarks and the chemical potential of the equilibrium system. Results with different $\alpha$ are shown in Fig. 1. It can be seen from Fig. 1 that no matter how much the $\alpha$ is, the number density of quarks is zero unless the chemical potential is greater than a value. This is a model-independent result proposed by Ref. \cite{30}. It is proved, based on a universal argument, that when the chemical potential $\mu$ is smaller than a value $\mu_0$, the quark-number density vanishes identically. Also, the curves of quark number densities break off at a critical point $\mu_c$ that suggests the chiral phase transition point. When $\alpha$ is 0.85, the number density curves become smooth because the phase transition becomes a crossover. Our value of $\alpha$ that changes the type of phase transition differs from our previous work \cite{20} slightly because Fig. 1 describes the beta-equilibrium matter where the chemical potentials of different kinds of quark are not the same. A very precise value of $\alpha$ should be obtained by calculating an order parameter such as the chiral susceptibility. But in this work, this value is not what we are interested in, because the jumps in the number density become smaller and smaller and the curves become continuous gradually as our $\alpha$ increases, and thus the stiffness of our EOS increases with $\alpha$, but no remarkable signature of where the transition becomes a crossover can be identified in the M-R relation.

\begin{figure}
   \begin{center}
      \includegraphics[width=0.5\textwidth]{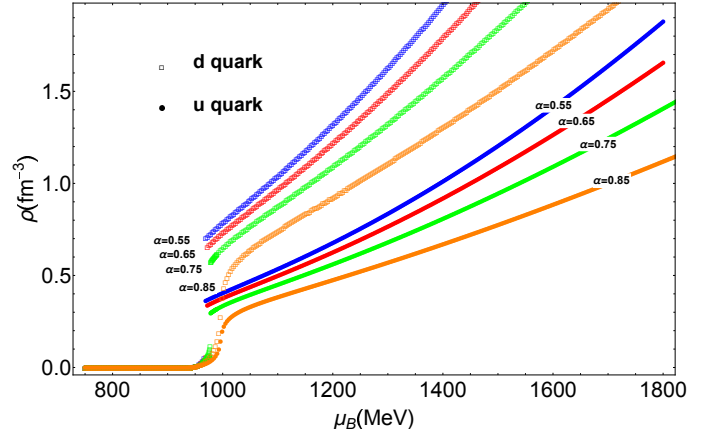} \caption{The densities of u,d quarks with $\alpha=0.55,0.65,0.75,0.85$, and $\mu _B$ is the baryon chemical potential. When $\alpha$ is 0.85, the number density curves become smooth because the phase transition has become a crossover.}
   \end{center}
\end{figure}
By definition, the he most general pressure formula of strong interaction matter at finite chemical potential and zero temperature is \cite{31,newzong}:

\begin{eqnarray}
\begin{aligned}
P\left(\mu _u, \mu _d;M \right) =& P\left(\mu =0;M \right) + \int_ 0^{\mu _u} \rho _u \left(\mu _u^{'},0 \right) \, d\mu _u^{'} \\
&+ \int_ 0^{\mu _d} \rho _d \left(\mu _u,\mu _d ^{'}\right) \, d\mu _d^{'}.
\end{aligned}
\end{eqnarray}
%Here, an integral constant is left to be determined. This is the vacuum  pressure at zero chemical potential.
where $P(\mu =0;M)$ denotes the vacuum pressure, which is a density-independent quantity and $M$ is a solution of the quark gap equation. Actually, the vacuum pressure can not be measured. The only one can be measured is the vacuum pressure difference and a typical example is the Casimir effect. To do this, we need to choose a reference ground state. This reference ground state should in principle be a trivial vacuum state of strong interaction system we are studying.  In the past, the trivial vacuum pressure is often marked as $P(0;m)$, here $m$ is the current quark mass. Like in the bag model, the  bag constant $B$ describes the pressure difference between the trivial vacuum  and non-trivial vacuum (Nambu vacuum, reflecting the spontaneous symmetry breaking of the vacuum), which is pointed out by Ref. \cite{17}. Thus, the bag constant  commonly used in the past is defined as
\begin{eqnarray}
	B=P\left(0,0 ; m, 0\right)-P(0,0 ; M, 0), \label{eq2}
\end{eqnarray}
where $M$ specially denotes the Nambu solution of the quark gap equation. It should be noted here, except in the chiral limit, $m$ is not a solution of the quark gap equations. Since $m$ is not a solution of the quark gap equation, it is not appropriate to select this state as a reference ground state.   In order to overcome the defect of Eq. (\ref{eq2}), the authors of Ref. \cite{xu} suggest that the quasi-Wigner vacuum (the  quasi-Wigner vacuum corresponding to quasi-Winger solution of the gap equation, details can be found in Refs. \cite{xu,add5,cui}) is regarded as the reference ground state and the bag constant is redefined as
%Beyond the chiral limit, formula (\ref{eq2}) is not well defined, because the $m$ is not a solution of the gap equations. Recently, a new algorithm is performed and the Wigner solution of gap equation can be found. Therefore, the bag constant can be defined as
\begin{eqnarray}
	B=P\left(0,0 ; M_{W}, 0\right)-P(0,0 ; M_{N}, 0), \label{eq3}
\end{eqnarray}
where, $M_{W}$ and $ M_{N}$ denote the quasi-Wigner and Nambu solution of the quark gap equation respectively. Therefore, the EOS that is actually used in our manuscript is
\begin{eqnarray}
P\left(\mu _u, \mu _d;M \right) =& -B + \int_ 0^{\mu _u} \rho _u \left(\mu _u^{'},0 \right) \, d\mu _u^{'} \\
&+ \int_ 0^{\mu _d} \rho _d \left(\mu _u,\mu _d ^{'}\right) \, d\mu _d^{'}..
\end{eqnarray}
 Here, it should be stressed that the EOS of the strong interaction matter depends on  not only the Nambu solution, but also the choice of the reference ground state, which is very important in the study on the structure of the neutron star.

Since the bag constant reflects the non-perturbative vacuum nature of QCD, it is difficult to calculate it from the first principle of QCD. In this case, we often use some non-perturbative QCD models for related calculations, such as  Dyson--Schwinger equations and NJL model. According to the new definition Eq. (\ref{eq3}), the bag constant calculated by the Dyson--Schwinger equations approach is $(171 \mathrm{MeV})^4$ \cite{xu}.  And in the framework of NJL model, different model parameters will give different results, as is shown clearly in Table. \ref{table}. The parameters utilized are calibrated by fitting hadronic experiment and LQCD data at zero temperature and zero chemical potential, such as the two-quark condensate, $m_{\pi}$ and $f_{\pi}$. The parameter set 2 is adopted in our work.
\begin{table}[htp!]
\centering
\caption{bag constant with different model parameters.}	\label{table}
\begin{tabular}{|c|c|c|c|c|}
\hline
set & $\Lambda$ [MeV] & $G \Lambda^2$ & $m$ [MeV]& B [MeV$^{4}$]\\ \hline
1   & 664.3           & 2.06          & 5.0      & (156)$^4$             \\ \hline
2   & 653.0           & 2.10          & 5.0      & (158)$^4$              \\ \hline
3  & 587.9           & 2.44          & 5.6      & (182)$^4$            \\ \hline
%3   & 569.3           & 2.81          & 5.5      & -234             \\ \hline
\end{tabular}
\end{table}
From  Table \ref{table}, it is easy to find that the bag constant depends on the model parameters, but it  agrees with the empirical domain \cite{33,34}, which ranges from $(100\mathrm{MeV})^4$ to $(200\mathrm{MeV})^4$ .  Since the bag constant plays an important role in the study on EOS of neutron star,  and at the same time we cannot calculate it from the first principle of QCD, in this case we treat the bag constant as a free parameter within the empirical domain, and see how it affects the stiffness of EOS. %It is usually associated with the bag constant B :$P\left(\mu =0 \right)=-B$. Because of the complications of the nonpertubative QCD vacuum, there is no reliable way to calculate the vacuum negative pressure. So it is considered as a free parameter to explore it's effect on the stiffness of our EOS. An empirical domain is $\left( 100~\mathrm{MeV} \right)^4-\left( 200~\mathrm{MeV} \right)^4$ according to Refs. \cite{23,33,34}.

 According to the thermodynamic relationship, the energy density of strong interaction matter is given by \cite{35,36}:

\begin{eqnarray}
\epsilon =-P + \sum _i \mu _i \rho _i.\label{eq4}
\end{eqnarray}
From Eq. (\ref{eq4}), it is easy to find that the  energy density $\epsilon$ depends on the pressure. As it is shown in Eq. (13), the pressure depends on the choice of the reference ground state. That is to say, whether  the minimal energy per baryon is discussed in the  quark degrees of freedom or the  hadron degrees of freedom,  a reference ground state must be chosen in advance.
%as the energy per baryon is discussed in the quark degree of freedom or hadron degree of freedom, a reference ground state  must be chosen in advance.
But what needs to be emphasized here is that we do not know how to do hadronization from the basic degrees of freedom of QCD, which means that we do not know the relationship between the reference ground of two different degrees of freedom. Therefore, we believe that it is not appropriate to judge whether the quark phase  or the hadron phase is more stable by using minimal energy per baryon as a criterion.
Therefore, inspired by Ref. \cite{28}, in this paper, we further assume that the two-flavor quark matter is more stable than the non-strange hadronic matter. Under this assumption, we would like to discuss if the current astronomical observations can rule out the possibility of non-strange quarks.

\section{Mass-radii relation of quark stars}
As is pointed out in Ref. \cite{20}, the chiral phase transition becomes a crossover if $\alpha$ is greater than 0.71, and $\mu_c =400~\mathrm {MeV}$ and 600 MeV is roughly corresponding to $\alpha=0.92$ and 1.04. So, we calculate the Mass-radii relations of quark stars with $\alpha=0.55, 0.65, 0.75, 0.85$ and $0.9$ respectively to show all possible situations. But the cutoff $\Lambda=653$ MeV we introduced into the NJL model indicates our model is valid under this energy scale only. So, $\alpha$ cannot be greater than 0.9 because the central chemical potential of d quark corresponding to the maximum mass of the neutron star must be smaller than our cutoff. The mass-radii relation can be obtained by solving the TOV equation:

\begin{eqnarray}
\begin{aligned}
\frac {dP\left(r \right)} {dr} &=-\frac {\left(\epsilon +P \right)\left(M + 4\pi r^3 P \right)} {r \left(r-2M \right)},\\
\frac {dM\left(r \right)} {dr} &=4\pi r^2 \epsilon.
\end{aligned}
\end{eqnarray}
To show the influence of $\alpha$ on the phase transition and stiffness of the EOS, several results with different values of $\alpha$ are compared in Fig. 2 while $B$ is fixed. Also, in Fig. 3, the constituent u quark mass is plotted to roughly indicate the transition of constituent quark masses.

\begin{figure}
   \begin{center}
      \includegraphics[width=0.5\textwidth]{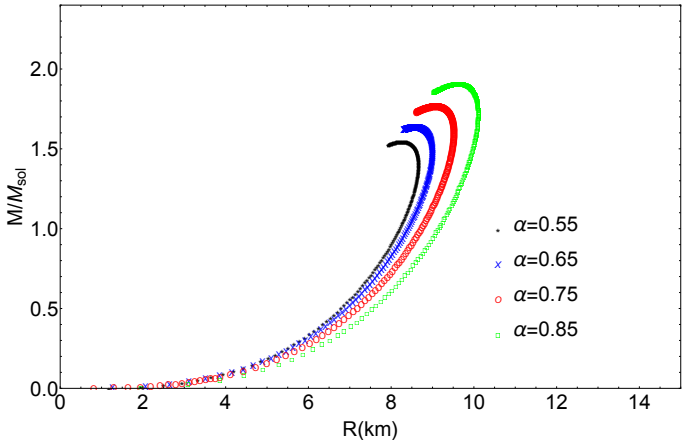} \caption{Mass-radii relations are exhibited with $\alpha=0.55,0.65,0.75,0.85$. $B=\left( 100 ~\mathrm{MeV} \right)^4$. The maximum mass increases with $\alpha$. Masses are scaled by the mass of sun: $M_{sol}$.}
   \end{center}
\end{figure}

\begin{figure}
   \begin{center}
      \includegraphics[width=0.5\textwidth]{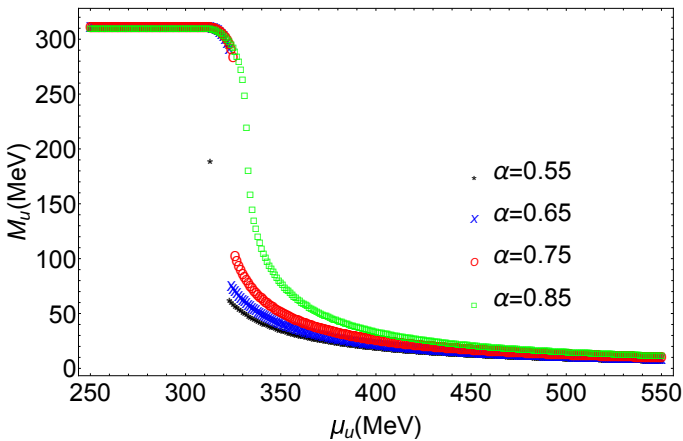} \caption{Constituent u quark mass as a function of u quark chemical potential. $\alpha=0.55,0.65,0.75,0.85$.}
   \end{center}
\end{figure}

As $\alpha$ increases, both $\mu_c$ and the stiffness of our EOS increase. They increase rapidly after $\alpha$ is greater than the point where  the chiral phase transition becomes a crossover. Just as what we have pointed out, if $\mu_c$ is small, dressed quarks will lose their dynamical mass, and the quark mass gradually become bare quark mass at an early stage because of the chiral restoration. This is in conflict with what we expected. However, in our modified NJL model, if $\alpha$ is greater, the point where constituent quarks lose their mass will be postponed, and the stiffness of our EOS will also increase. To explore the effects of the negative pressure of vacuum, mass-radii relations with different values of B are exhibited in Fig. 4.

\begin{figure}
   \begin{center}
      \includegraphics[width=0.5\textwidth]{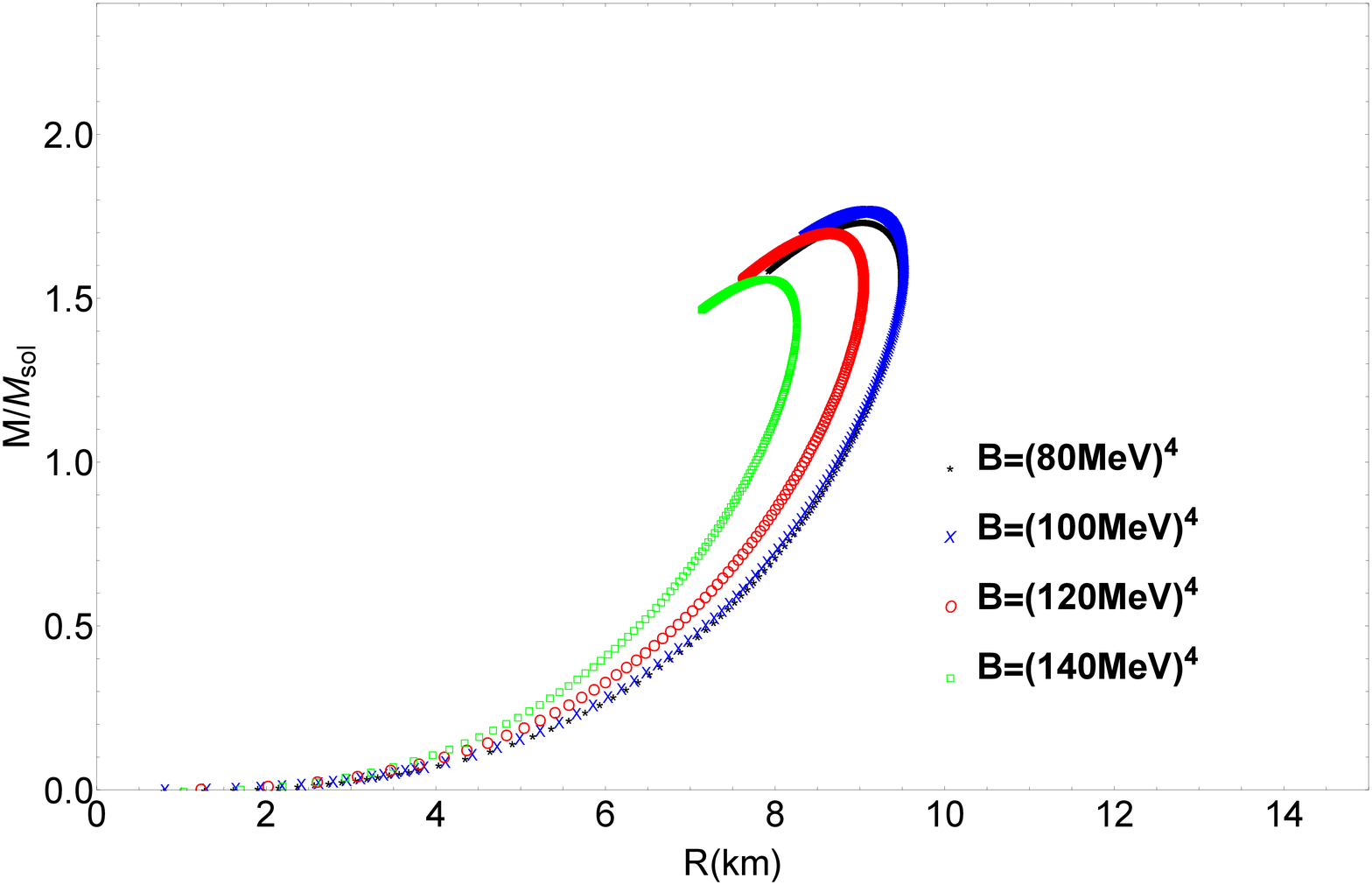} \caption{Mass-radii relations are exhibited with $B=\left(80~\mathrm {MeV} \right)^4, \left(100~\mathrm {MeV} \right)^4, \left(120~\mathrm {MeV} \right)^4, \left(140~\mathrm {MeV} \right)^4$ and $\alpha=0.75$. Here the maximum mass is $1.75 M_{sol}$.}
   \end{center}
\end{figure}

From Fig. 4 we can see that the maximum mass of the quark star doesn't increase with $B$ monotonically. It reaches a maximum value at $B=\left(100~\mathrm {MeV} \right)^4$. Of course, the corresponding value of $B$ will also change with $\alpha$. So, by manipulating the combination of $B$ and $\alpha$ slightly, we can promote the stiffness of our EOS greatly and construct a two-solar-mass quark star. This is because the contribution from the vector term is retained by the Fierz transformation, which is in accordance with the result in Ref. \cite{compare}. \\

\begin{figure}
   \begin{center}
      \includegraphics[width=0.5\textwidth]{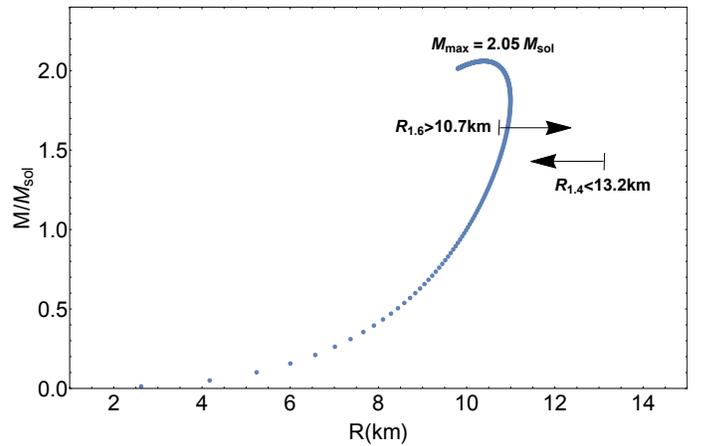} \caption{A mass-radius relation where the maximum mass can reach $2.05~M_{sol}$. $\alpha=0.9$, $B= \left(90~\mathrm {MeV} \right)^4$. This result is compared with the radius constraint from observations.}
   \end{center}
\end{figure}
In a recent observation, a super massive neutron star with about 2.17 times of solar mass at $68.3\%$ credibility is reported in Ref. \cite{mrnew}. This value is also the upper limit on neutron star mass given by a study combining electromagnetic and gravitational wave information on the binary neutron star merger GW170817 \cite{max}. The maximal quark star mass computed in our model is greater than two times of solar mass but smaller than this value. In Fig. 5, the mass-radius relation is plotted with $\alpha=0.9$ and $B=\left(90~\mathrm {MeV} \right)^4$. Important data are included in Table II.
\begin{table*}[htp!]
\centering
\caption{properties of our two-solar-mass quark star. }	\label{tab}
\begin{tabular}{|c|c|c|c|c|c|c|c|c|}
\hline
$\alpha$  & B    & $M_{max}$ & $\rho_{min}$ & $\rho_c$ & $\rho_{cutoff}$&$R_{M=max}$&$R_{1.6}$&$R_{1.4}$ \\ \hline
0.9   & $\left(90~\mathrm {MeV} \right)^4$   & 2.05 $M_{sol}$   & 0.157 $fm^{-3}$  & 0.802 $fm^{-3}$  & 0.808 $fm^{-3}$&10.5 km&10.9km&10.8km         \\ \hline
\end{tabular}
\end{table*}
 
In the Table II, $M_{max}$ is the maximum quark star mass, the corresponding radius is $R_{M=max}$, and $\rho_{min}$ donates the minimal baryonic density on the surface of the star. The finite density at the surface of the quark star is due to the bag constant we introduce. $\rho_c$ is the central desity corresponding to the maximum mass and $\rho_{cutoff}$ is the baryonic density for which the Fermi momentum of the d quark is equal to the cutoff.$R_{1.6}$ and $R_{1.4}$ are the radii corresponding to the 1.6-solar-mass quark star and 1.4-solar-mass quark star. Our maximal mass is 2.05 $M_{sol}$, and the central baryonic density corresponding to it is 0.802 $fm^{-3}$ that is slightly smaller than the baryonic density for which the Fermi momentum of the d quarks is equal to the cutoff. Besides, when the rotation of the quark star is taken into consideration, the value of the maximum mass is roughly $10\%-20\%$ higher than the non-rotating case \cite{rot}.

\section{Summary and discussion}
As astronomy observations are accumulating, a reliable EOS to describe the cold condense matter in compact stars is desirable to astronomers. In our model, a parameter $\alpha$ used to reflect the weight of different interaction channels. This parameter can influence the nature of EOS greatly. In this paper, different results with different $\alpha$ are compared. If $\alpha$ is 0.5, our model reduces to the normal mean field approximation model in Ref. \cite{29}. However in this work we find that as $\alpha$ increases, the critical chemical potential $\mu_c$ increases, the chiral phase transition becomes a crossover and the stiffness of EOS can increase greatly to support a two-solar-mass pure quark star. This cannot be realized if $\alpha$ is treated naively as a constant 0.5. So, our conclusions are: 1. The stiffness of our EOS for quark matter will be influenced greatly if the critical chemical potential is increased when a greater $\alpha$ is introduced. 2. The existence of two-flavor quark stars cannot be ruled out from the observation of neutron star maximum mass.

Finally some prospects for further studies are given here: First, it should be pointed out that in our modified NJL model, the three-momentum cutoff is applied to all the integrals. So, to get a reliable result, the chemical potential of quarks cannot be greater than the cutoff. This means the chemical potential in the center of the quark star cannot be too high, and because of this we abandon the result with an $\alpha$ greater than 0.9. But this constraint will disappear if a better regularization scheme is introduced in our model. Second, only two-flavor quark stars are discussed in this paper while strange quark matter EOS can also be obtained by performing the same routine, which can be compared with this model. Third, Although, in this paper, only the maximum mass of neutron stars is discussed because it is the most reliable and accurate data that can be extracted from astronomy observations. Other properties such as the tidal deformability from GW170817 can also be utilized to constrain the parameter space in our model. But it can also be translated to the constraint on the radius. For example, the upper limit on the tidal deformability is often associated with the upper limit on the radius of a 1.4-solar-mass neutron star. The results from recent three works are $R \leq 13.76~km$, $R \leq 13.6~km$, and $8.9~km \leq R \leq 13.76~km$ respectively \cite{r1,r2,r3}. And the lower limit on the radius imposed on a 1.6-solar-mass neutron star is $10.7~km$ \cite{8}. When $\alpha$ is 0.9 and $B=\left(90~\mathrm {MeV} \right)^4$, the maximum mass of our pure quark star is $2.05~M_{sol}$, the radius of a 1.4-solar-mass quark star is $10.8~km$, and the radius of a 1.6-solar-mass quark star is $10.9~km$. This result meets all the observations above quite well, which is shown in Fig. 5. To explore the range of our parameters $\alpha$ and B that satisfy the observations, some results with different parameters are listed in Table III.

\begin{table*}[htp!]
\centering
\caption{Results with different parameters. }	\label{tab1}
\begin{tabular}{|c|c|c|c|c|}
\hline
$\alpha$  & B    & $M_{max}$ &$R_{1.6}$&$R_{1.4}$ \\ \hline
0.9   & $\left(100~\mathrm {MeV} \right)^4$   & 2.01 $M_{sol}$&10.5 km&10.2 km         \\ \hline
0.9   & $\left(90~\mathrm {MeV} \right)^4$   & 2.05 $M_{sol}$&10.9 km&10.8 km         \\ \hline
0.9   & $\left(80~\mathrm {MeV} \right)^4$   & 2.11 $M_{sol}$&11.5 km&11.3 km         \\ \hline
0.8   & $\left(80~\mathrm {MeV} \right)^4$   & 2.00 $M_{sol}$&10.9 km&10.7 km         \\ \hline
\end{tabular}
\end{table*}

Besides the observed gravitational-wave signal from the neutron star merger GW170817, further gravitational-wave data will be available in the near future. In the recent general-relativistic simulation of merging neutron stars including quarks at finite temperatures, the authors in Ref. \cite{37} point out that the post-merger gravitational-wave spectrum can identify the phase transition from hadronic matter to quark matter during the process of the neutron star merger. So, Further constraints from both experiments and observations are necessary.

\acknowledgments
This work is supported in part by the National Natural Science Foundation of China (under Grants No. 11475085, No. 11535005, No. 11690030, and No.11873030), the Strategic Priority Research Program of the Chinese Academy of Sciences Multi-waveband Gravitational Wave Universe (Grant No. XDB23040000), the National Major state Basic Research and Development of China (Grant No. 2016YFE0129300).

\end{document}